\documentclass[11pt,
onecolumn,
 aip,
 jmp,%
 showkeys,
 amsmath,amssymb,
preprint,%
]{revtex4-1}

\usepackage[breaklinks=true,colorlinks=true,linkcolor=blue,urlcolor=blue,citecolor=blue]{hyperref}
\usepackage{graphicx}%
\usepackage{graphics}%
\usepackage{epstopdf}
\usepackage{dcolumn}%
\usepackage{soul,color}
\usepackage{bm}% bold math
\usepackage[mathlines]{lineno}% Enable numbering of text and display math
\expandafter\ifx\csname package@font\endcsname\relax\else
 \expandafter\expandafter
 \expandafter\usepackage
 \expandafter\expandafter
 \expandafter{\csname package@font\endcsname}%
\fi
\begin{document}
\abovedisplayskip=3pt
\belowdisplayskip=3pt
\abovedisplayshortskip=2pt
\belowdisplayshortskip=2pt

%\preprint{AIP/123-QED}

\title{\Large{Radiofrequency generation by coherently moving fluxons}}
\author{O. V.~Dobrovolskiy}
    \email{Dobrovolskiy@Physik.uni-frankfurt.de}
    \affiliation{Physikalisches Institut, Goethe University, 60438 Frankfurt am Main, Germany}
    \affiliation{Physics Department, V. Karazin National University, 61077 Kharkiv, Ukraine}
\author{R. Sachser}
\author{M. Huth}
    \affiliation{Physikalisches Institut, Goethe University, 60438 Frankfurt am Main, Germany}
\author{V. A. Shklovskij}
    \affiliation{Physics Department, V. Karazin National University, 61077 Kharkiv, Ukraine}
\author{R.~V.~Vovk}
    \affiliation{Physics Department, V. Karazin National University, 61077 Kharkiv, Ukraine}
    \affiliation{ICST Faculty, Ukrainian State University of Railway Transport, 61050 Kharkiv, Ukraine}
\author{V.~M.~Bevz}
    \affiliation{ICST Faculty, Ukrainian State University of Railway Transport, 61050 Kharkiv, Ukraine}
\author{M.~I.~Tsindlekht}
    \affiliation{The Racah Institute of Physics, The Hebrew University of Jerusalem, 91904 Jerusalem, Israel}
\date{\today}

\begin{abstract}
A lattice of Abrikosov vortices in type II superconductors is characterized by a periodic modulation of the magnetic induction perpendicular to the applied magnetic field. For a coherent vortex motion under the action of a transport current, the magnetic induction at a given point of the sample varies in time with a washboard frequency $f_\mathrm{WB} = v/d$, where $v$ is the vortex velocity and $d$ is the distance between the vortices in the direction of motion. Here, by using a spectrum analyzer connected to a 50\,nm-wide Au nanowire meander near the surface of a superconducting Nb film we detect an ac voltage induced by coherently moving fluxons. The voltage is peaked at the washboard frequency, $f_\mathrm{WB}$, and its subharmonics, $f_\mathrm{TOF} =  f_\mathrm{WB}/5$, determined by the antenna width. By sweeping the dc current value, we reveal that $f_\mathrm{WB}$ can be tuned from $100$\,MHz to $1.5$\,GHz, thereby demonstrating that patterned normal metal/superconductor nanostructures can be used as dc-tunable generators operating in the radiofrequency range.
\end{abstract}

%\pacs{74.45.+c, 78.67.Uh, 72.15.Eb}

\maketitle

Magnetic fields are known to penetrate type II superconductors in the form of Abrikosov vortices\cite{Abr57etp}, or fluxons, each carrying one quantum of magnetic flux $\Phi_0 = 2.07\times10^{-15}$\,Vs. The vortices form a triangular lattice and can be regarded as tiny solenoids producing local magnetic field variations extending over distances of $2\lambda$ perpendicular to the applied field, where $\lambda$ is the magnetic penetration depth. The maximum field, $B_{v}\approx {\frac {\Phi_{0}}{2\pi \lambda ^{2}}}(\ln \lambda/\xi)$ \cite{Bra95rpp}, occurs in the vortex core with a diameter of $\simeq2\xi$, where $\xi$ is the superconducting coherence length. A transport current of density $\mathbf{j}$ exerts a Lorentz force $\mathbf{F}_L = \mathbf{j} \times \mathbf{B}$ on the flux lattice. If $\mathbf{F}_L$ exceeds the pinning force, the vortex lattice moves and the oscillations of the magnetic induction at a given point in space $\mathbf{B}(x,y,t)$ are characterized by the washboard frequency $f_\mathrm{WB} = v/d$, where $v$ is the vortex velocity and $d$ is the distance between the vortices in the direction of motion.

Previously, the washboard frequency of the moving vortex lattice was detected by ac/dc interference \cite{Har95prl} in high-temperature superconducting cuprates. The detected sharp frequencies \cite{Har95prl} were \emph{intrinsic} to the moving lattice whose uniform motion in a random distribution of pinning sites produced a periodic modulation of the pinning force. If the dc driving current is augmented by an rf component an interference between the intrinsic oscillations and the superimposed ac current occurs when both frequencies are harmonically related \cite{Fio71prl}. Quantum interference effects are seen as Shapiro steps in the current-voltage characteristics \cite{Sha63prl,Mar75ssc,Dob17pcs} when one or a multiple of the hopping period of Abrikosov vortices in superconductors with periodic pinning coincides with the period of the ac drive. In a different physical context, voltage oscillations at the washboard frequency were observed as narrow band noise in the sliding state of charge-density wave conductors \cite{Fle79prl}, as well as in sliding spin-wave states \cite{Bar93prl}.

The different dynamic states and noise associated with the vortex motion in type II superconductors was studied in several numerical works \cite{Ols98prl,Ols98prl2}. In particular, they predicted the broad band noise near depinning, the time of flight signal and the washboard signal as well as the signal shift towards higher frequencies with increasing dc drive. Recently, a similar study was also done for skyrmion lattices \cite{Dia17prb}. The spectral analysis of the skyrmion velocity noise fluctuations has revealed broad band, time of flight and narrow band noise signals as well.

As distinct from the previously mentioned ``intrinsic'' effects, according to Faraday's law, a time-varying magnetic flux within a closed loop of wire produces an electromotive force (emf) or voltage (within a circuit). This makes it feasible to observe the rf signal associated with the vortex motion \emph{externally}, by placing a device which is sensitive to the magnetic flux near the surface of a superconductor. Currently, the highest flux sensitivity is provided by SQUIDs \cite{Fag06rsi} which allow one to detect signals produced by the spin magnetic moment of a single electron \cite{Vas13nat}. SQUID-based devices are successfully used for the visualization of nanoscale inhomogeneous magnetic states \cite{Ana16nac} and the high-velocity dynamics of individual vortices \cite{Emb17nac}. In general, electromagnetic radiation from a moving vortex lattice is also expected as it comes to a sample surface \cite{Bul06prl}. This electromagnetic radiation at $f_{WB}$  and its harmonics up to the superconducting gap frequency is generated by oscillating electric and magnetic fields of vortices near the surface and propagates into free space due to the continuity of tangential components of the fields at the surface. The effect of an external high-frequency stimulus on the dynamics of Abrikosov vortices in superconducting thin films has also been investigated experimentally \cite{Lar15nsr,Dob15met,Lar17pra,Dob17nsr} and theoretically \cite{Pom08prb,Shk11prb}. In particular, experiments provided evidence for stimulation of superconductivity by microwaves due to the presence of vortices \cite{Lar15nsr} and for enhanced stability of thermally-induced vortex avalanches under a microwave field excitation \cite{Lar17pra}.

In the particular case of coherently moving fluxons their dynamics can also be studied using hybrid, superconducting/normal metal detector structures. Given that nanowires with sub-100\,nm width can be fabricated by modern nanofabrication techniques, vortex velocities of the order of 100\,m/s allow one to achieve ac voltage frequencies, associated with crossing of nanowires by vortices, on the order of 1\,GHz. Here, we study the frequency-domain voltage response of an Au meander nanowire which picks up the magnetic induction of magnetic flux quanta in an adjacent superconducting Nb thin film. While the meander layout is similar to those widely used in superconducting nanowire single-photon detectors (SNSPDs) \cite{Rat15lsa}, here we use a quintuple, normally conducting meander with an active area of $1\times2.5$\,$\mu$m$^2$. The planar layout of the generator, designed to operate in a perpendicular magnetic field of $45$\,mT, allows for easy on-chip integration with other fluxonic devices, such as diodes \cite{Vil03sci}, microwave filters \cite{Dob15apl} and transistors \cite{Vla16nsr}.

\begin{figure}[t!]
    \centering
    \includegraphics[width=0.75\linewidth]{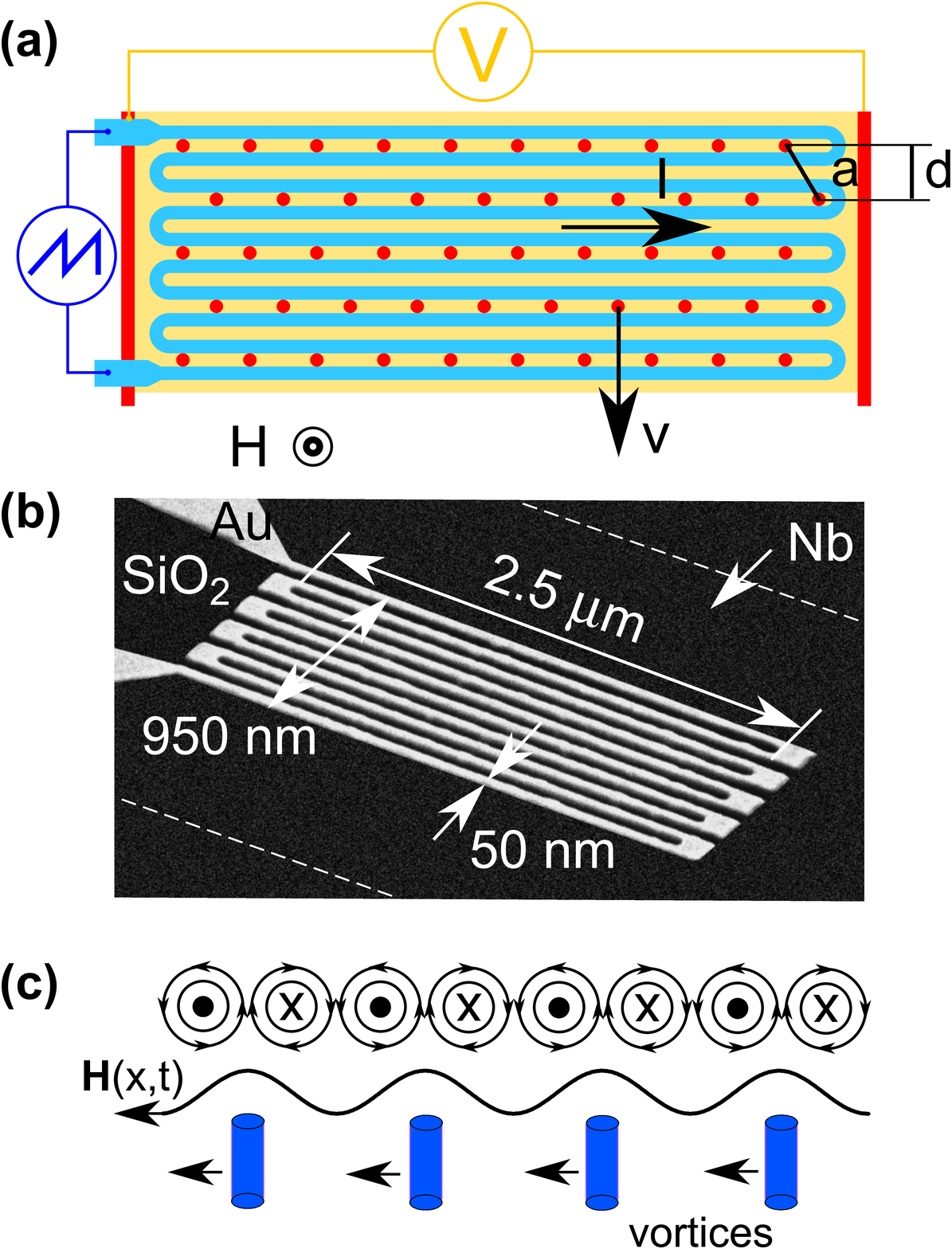}
    \caption{(a) Experimental geometry (not to scale). A nanowire antenna is placed on top of the Nb bridge. The transport current $\mathbf{I}$ induces a Lorentz force on the vortices which move across the meander with velocity $\mathbf{v}$. The vortices are depicted as red points between the meander loops. The arrangement of vortices corresponds to the matching field $H = 45$\,mT, where $a$ is the intervortex distance and  $d$ is the distance between the vortex rows in the direction of motion. (b) Scanning electron microscopy image of the meander nanowire. The Nb bridge is mounted face-to-face to the nanowire within the area indicated by the white dash. (c) The magnetic flux emanating from the moving vortices induces an rf magnetic field wich is picked up by the nanowire. The associated rf voltage is registered by a spectrum analyzer connected to the antenna.}
    \label{f1}
\end{figure}

The device concept is sketched in Fig. \ref{f1}(a). The device is based on a 45\,nm-thick, 50\,nm-wide Au meander nanowire connected to a high-frequency oscilloscope/spectrum analyzer (Tektronix MDO4024C) operating in the frequency domain. The nanowire was fabricated from an Au film by focused ion beam milling in a high-resolution scanning electron microscope (FEI Nova NanoLab 600). The Au film was prepared by dc magnetron sputtering onto Si/SiO$_2$ substrate with a pre-sputtered 5\,nm-thick Cr buffer layer. The thickness of the SiO$_2$ layer was 200\,nm. In the sputtering process the substrate temperature was $T = 22^\circ$C, the growth rate was $0.055$\,nm/s and $0.25$\,nm/s, and the Ar pressure was $2\times10^{-3}$\,mbar and $7\times10^{-3}$\,mbar for the Cr an Au layers, respectively. The nanowire consists of five periods (loops) to pickup the magnetic induction of moving vortices. A scanning electron microscopy image of the as-fabricated meander is shown in Fig. \ref{f1}(b). The active area of the meander antenna was covered with a 10\,nm-thick Al$_2$O$_3$ layer to avoid direct electrical contact with the superconductor which would otherwise shunt it. A four-point bridge was defined by UV-lithography in a 50\,nm-thick Nb film with a width of $3$\,$\mu$m and a distance between the voltage contacts of $L = 10$\,$\mu$m. The epitaxial (110) Nb film was sputtered by dc magnetron sputtering onto an a-cut sapphire substrate. In the sputtering process the film growth rate was $0.5$\,nm/s, the substrate temperature was $T = 850^\circ$C, and the Ar pressure was $5\times10^{-3}$\,mbar. The Nb film has a superconducting transition temperature of $9.05$\,K and a room-to-10\,K resistance ratio of 27. The upper critical field of the Nb film at zero temperature $H_{c2}(0)$ is $750$\,mT as deduced from fitting the dependence $H_{c2}(T)$ to the phenomenological law $H_{c2}(T) = H_{c2}(0)[1 -(T/T_c)^{1/2}]$. The values of the superconducting coherence length $\xi(0)$ deduced from $H_{c2}(0)$ were found to be around $21$\,nm. The estimated magnetic field penetration depth $\lambda(0)$ in the Nb film amounts to $100$\,nm \cite{Gub05prb}.

The Nb bridge was positioned by microscrews in the area encaged by the white dashed lines in Fig. \ref{f1}(b). The in-plane angle alignment error with respect to the nanowire edges was less than $0.2^\circ$. The Au meander was positioned in the geometrical center of the Nb bridge, to diminish the contribution of vortices pinned at the bridge edges. The measurements were done in a magnetic field $H = 45$\,mT directed perpendicular to the film surface. A dc transport current was applied through the Nb film parallel to the edges of the meander thus causing the vortices to move in the perpendicular direction, Fig. \ref{f1}(a). The nanowire was connected via a $50\,\Omega$-matched feed line to the port of the oscilloscope and we measured the rf voltage amplitude induced in the nanowire in terms of the spectral power density $A$ in the frequency range 50\,MHz to 3\,GHz.

\begin{figure}[t!]
    \centering
    \includegraphics[width=1\linewidth]{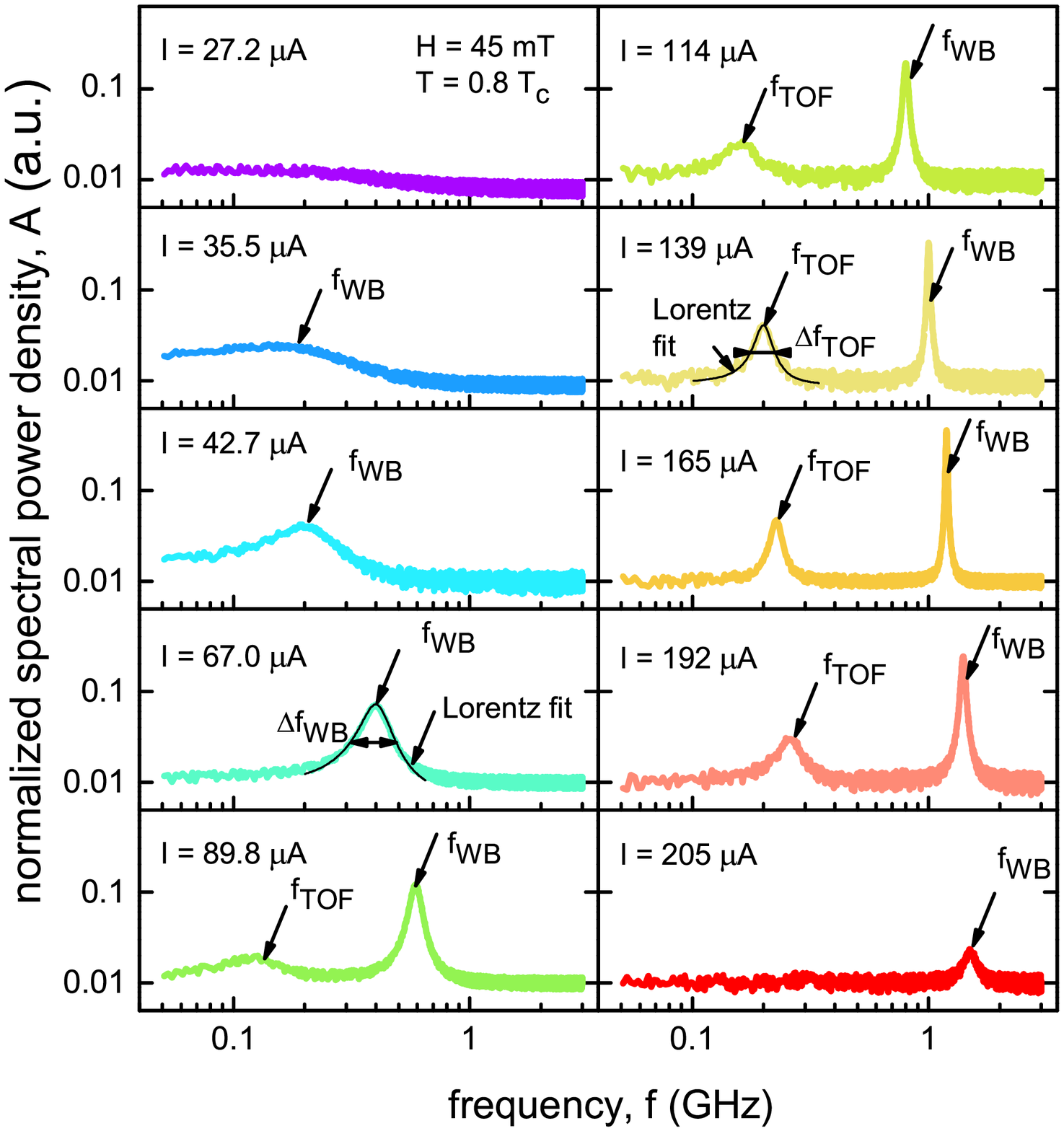}
    \caption{Frequency spectra recorded by the spectrum analyzer for a series of dc currents flowing through the Nb film at $T = 0.8T_c$ and $H = 45$\,mT. The arrows indicate the central frequencies $f_\mathrm{WB}$ and $f_\mathrm{TOF}$ at which the measured signal is peaked. The black solid lines are fits to the Lorentz distribution exemplifying the respective linewidths, $\Delta f_\mathrm{WB}$ and $\Delta f_\mathrm{TOF}$.}
    \label{f2}
\end{figure}

Figure \ref{f2} displays the frequency traces recorded by the spectrum analyzer for a series of dc currents flowing through the Nb film. At $I=0$, there is no signal registered by the analyzer down to the noise floor at the -60\,dBmV level. At small currents $I \lesssim 20\,\mu$A the traces are flat, whereas at $I\gtrsim 35\,\mu$A a broad peak at $f_\mathrm{WBmin}\simeq180$\,MHz appears. With an increase of $I$ from $35\,\mu$A to $165\,\mu$A the magnitude of the peak increases and its frequency shifts to $1.2$\,GHz. This is accompanied by the appearance of a second, broader and smaller peak on the low-frequency side of the large peak, which becomes visible at $I \gtrsim 85\,\mu$A at $f_\mathrm{TOFmin}\simeq100$\,MHz. The maximal amplitude and the minimal linewidth of both peaks is observed at $I \thickapprox 165\,\mu$A at $f_\mathrm{TOF} = 240$\,MHz and $f_\mathrm{WB} = 1.2$\,GHz, respectively. With a further increase of the current, the peaks become broader and their magnitudes decrease. The left peak vanishes at $f_\mathrm{TOFmax} \backsimeq300$\,MHz at $I = 200\,\mu$A while the right peak remains visible up to $f_\mathrm{WBmax} \backsimeq 1.5$\,GHz at $I = 210\,\mu$A.

\begin{figure}[t!]
    \centering
    \includegraphics[width=1\linewidth]{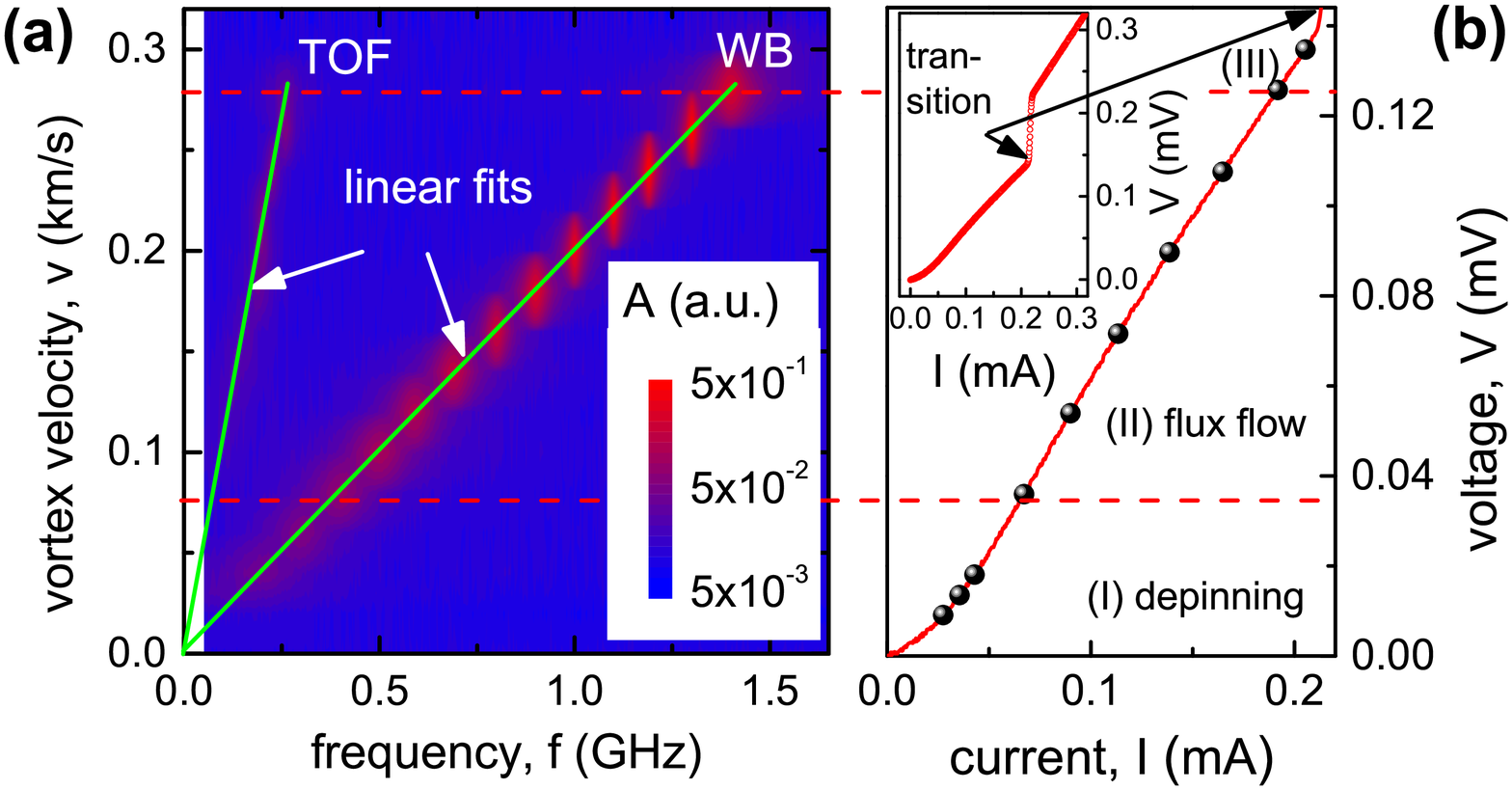}
    \caption{(a) Normalized spectral power density as a function of frequency and vortex velocity at $T=0.8T_c$ and $H = 45$\,mT. The washboard (WB) mode and the time-of-flight (TOF) mode are indicated. The solid lines are linear fits. (b) The current-voltage curve (CVC) of the Nb film at $T=0.8T_c$ and $H = 45$\,mT. The circles in the CVC correspond to the current values in Fig. \ref{f2}. The transition of the sample into the normal state at larger currents is shown in the inset. The horizontal dashed lines demarcate the different regimes in the CVC and the associated rf voltage response.}
    \label{f3}
\end{figure}

For a deeper analysis of the peaks associated with the vortex dynamics, in Fig. \ref{f3} we present the normalized spectral power density as a function of frequency and vortex velocity in comparison with the current-voltage-curve (CVC) of the Nb film. The vortex velocity was deduced from the CVC by the relation $v = V/BL$, where $V$ is the measured voltage, $B=45$\,mT is the applied magnetic field, and $L = 10\,\mu$m is the distance between the voltage contacts. In the contour plot in Fig. \ref{f3}(a), two bright areas associated with the peaks in the spectra in Fig. \ref{f2} can be recognized. The broader and brighter one, which almost diagonally extends through the entire plot, corresponds to $f_\mathrm{WB}(v_\mathrm{WB})$. The darker one, best seen between $0.2$\,GHz and $0.3$\,GHz at higher vortex velocities, relates to $f_\mathrm{TOF}(v_\mathrm{TOF})$. The variation of color along the bright areas is associated with the coarse-grained vortex velocity points deduced from the CVC and included into the contour plot.

To elucidate what geometrical parameters of our system link the peak frequencies with the vortex velocities, we fit the bright areas in Fig. \ref{f3}(a) to straight lines $v_\mathrm{TOF} = C_\mathrm{TOF} f_\mathrm{TOF}$ and $v_\mathrm{WB} = C_\mathrm{WB} f_\mathrm{WB}$, varying $C_\mathrm{TOF}$ and $C_\mathrm{WB}$ as fitting parameters. The best fits are obtained with $C_\mathrm{WB} = 203\pm6$\,nm and $C_\mathrm{TOF} = 1.05\pm0.03\,\mu$m as shown in Fig. \ref{f3}(a). $C_\mathrm{WB}$ corresponds very well to the meander period $d = 200\,$nm. $C_\mathrm{TOF}$ is slightly larger than the distance between the outer edges of the meander nanowire $W =950\,$nm. If we assume a triangular flux lattice with the parameter $a = (2\Phi_0/\sqrt{3}H)^{1/2} = 231$\,nm at $H = 45$\,mT and the matching condition $d = a\sqrt{3}/2$, the arrangement of vortices in Fig. \ref{f1}(a) allows us to explain both observed peaks as follows. At $45$\,mT the triangular vortex lattice is commensurate with the meander nanowire and all vortices contribute to the produced voltage in phase. In this case the distribution of the magnetic induction is periodic in space $\mathbf{B}(t,x,y) \equiv \mathbf{B}(t,x+na,y+ma)$ and in time, where $n$ and $m$ are integers. The out-of-plane field components emanating from the vortices induce an Oersted field around the nanowire. As the vortices move, Fig. \ref{f1}(c), the Oersted components oscillate in time and thereby produce an alternating ac voltage. Accordingly, the peak at $f_\mathrm{WB}$ occurs due to the vortices overcoming one period of the meander. The difference between $C_\mathrm{TOF}$ and $W = 950$\,nm can be understood if we take into account two additional vortex rows at the outer edges of the meander, which contribute to the measured signal. If we add the radii of the vortices $\sim \xi(0.8T_c) \approx 27$\,nm on each side of the meander to the meander width, the resulting $W+2\xi\backsimeq1\,\mu$m is in good agreement with $C_\mathrm{TOF}$. Therefore, the broader, low-frequency peak is related to the time needed for a vortex row to cross the meander from one outer edge to the other one. When the field is tuned in the range $44\,$mT $<H< 50$\,mT away from the matching condition, the peaks become broader but remain visible. At magnetic field values outside of this range no peaks are observed, which can be explained by the lost long-range order in the vortex lattice with respect to the meander. Therefore, we conclude that the coherence of the vortex dynamics is decisive for the observation of the rf generation by vortices.

\begin{figure}[t!]
    \centering
    \includegraphics[width=1\linewidth]{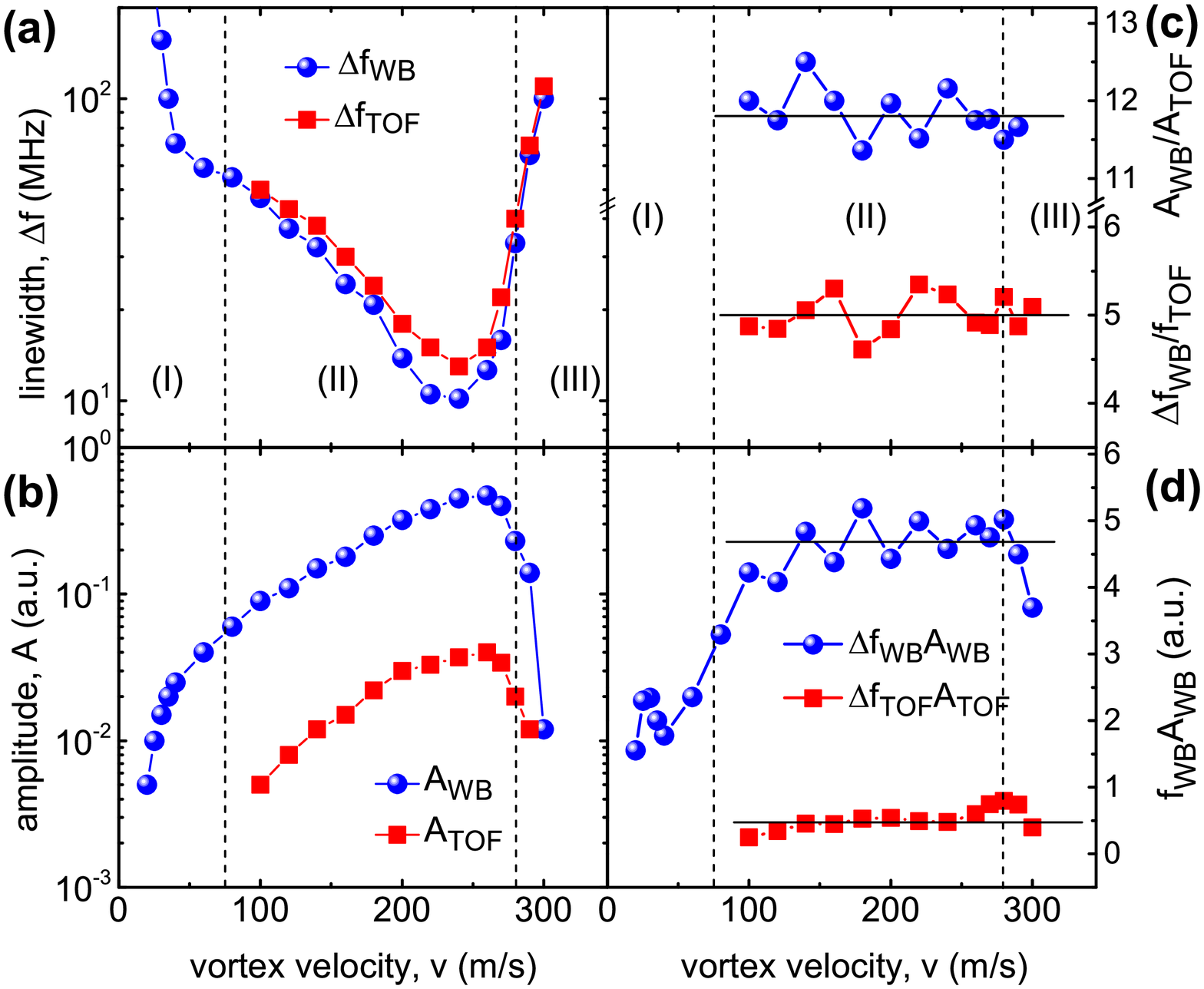}
    \caption{Linewidths (a) and peak amplitudes (b) versus vortex velocity. The areas under the peaks (c) and the relative ratios of the linewidths and peak amplitudes (d) versus vortex velocity. Regions (I), (II) and (III) correspond to those in Fig. \ref{f3}(b).}
    \label{f4}
\end{figure}

The rf generation by vortices has been observed between $0.5T_c$ and $0.8T_c$. At $T \geq 0.85T_c$ the amplitude of the detected signal decreased below the sensitivity level of our setup. This can be explained by the increase of the magnetic penetration depth $\lambda$ at $T\rightarrow T_c$ such that the modulation of the magnetic induction at the vortex cores and between them becomes very small. The particular set of data reported here was acquired at $T = 0.8T_c$ since at this temperature the depinning current is yet rather small and the peak frequency can be efficiently tuned by rather small currents. With $\lambda(0.8T_c) \approx 130$\,nm and $\xi(0.8T_c) \approx 27$\,nm the magnetic induction variation at the vortex core and between the vortices can be estimated as $\Delta B = B_v - B_{min} \sim 10$\,mT, where $B_{v} \approx 32$\,nm, thus yielding $\Delta B / B_v \approx 30\%$ \cite{Bra95rpp}. According to Faraday's law, a uniform $30\%$-change of the magnetic flux for a single vortex moving with velocity $v= 100$\,m/s should produce an emf of $\varepsilon = - \Delta \Phi/\Delta t = - 0.3\frac{v\Phi_0}{d/2} \sim 0.6\,\mu$V. Assuming that coherently moving vortices contribute to the emf additively,  the emf produced by 50 vortices in our sample can be estimated as $\varepsilon \sim 30\mu$V. Obviously, this simple estimate is an upper-bound estimate as it does not account for signal losses due to the impedance mismatch between the nanowire and feed line.

We now proceed to a quantitative analysis of the observed peaks. Both peaks can be fitted to the Lorentz distribution, which allows for a systematic analysis of their linewidth, $\Delta f$, and magnitude, $A$, as a function of the vortex velocity. The deduced dependences $\Delta f_\mathrm{WB}(v)$ and $\Delta f_\mathrm{TOF}(v)$ are displayed in Fig. \ref{f4}(a). The linewidth of both peaks exhibits a minimum at $v \thickapprox250$\,m/s with $\Delta f \backsimeq10$\,MHz. The amplitude of both peaks attains a maximum at the same velocity, Fig. \ref{f4}(b). The ratio $f_\mathrm{WB}/f_\mathrm{TOF}$ remains almost constant and equal to $5$ for vortex velocities between $100$\,m/s and $280$\,m/s, Fig. \ref{f4}(c). The ratio $A_\mathrm{WB}/A_\mathrm{TOF} \approx12$ is also almost constant in this range of vortex velocities. The fact that the magnitude of the peak at $f_\mathrm{WB}$ is an order of magnitude larger than that of the peak at $f_\mathrm{TOF}$ can be explained by the five periods of the meander antenna acting in phase. If we introduce the products $A_\mathrm{WB}\Delta f_\mathrm{WB}$ and $A_\mathrm{TOF}\Delta f_\mathrm{TOF}$ as measures for the areas under the peaks, these will allow us to analyze the number of vortices contributing coherently to the measured rf voltage. Interestingly, at vortex velocities between $100$\,m/s and $280$\,m/s these products are almost constant, Fig. \ref{f4}(d), suggesting that the number of vortices contributing to the signal is constant and we deal with a coherent vortex dynamics.

Outside of this range of vortex velocities, the area under the peak decreases, indicating that the number of vortices which takes part in the in-phase induction of voltage is decreasing as well. At small velocities this can be understood as a consequence of the depinning transition in the vortex dynamics. Indeed, even in high-quality films there is always a variation in the individual pinning forces acting on different vortices so that the long-range order in the vortex lattice is lost below the depinning transition. This corresponds to regime (I) in the CVC in Fig. \ref{f3}(b). The spatial order of the vortex lattice in relation to the voltage noise spectra was studied numerically in Ref. \cite{Ols98prl}. It has been shown that the vortex velocity distribution function is most broad in the plastic phase near depinning and it is most narrow in the crystallinelike phase, just as we observe in our experiment. Namely, as soon as the viscous regime of flux flow is established, regime (II) in Fig. \ref{f3}(b), the long-range order in the vortex lattice is recovered with increasing vortex velocity. This corresponds to a dynamic crystallization \cite{Kos94prl,Yar94prl} of the vortex lattice and this regime is stable as long as no further non-linearity comes into play. At high vortex velocities, regime (III), the electric field caused by the vortex motion accelerates quasiparticles in the vortex cores, which may escape from them. At $T\lesssim T_c$ the escape of quasiparticles leads to a shrinkage of the vortex core that, in return, causes a reduction of the vortex viscosity. The associated reduction of the viscous force impeding the vortex motion leads to a further increase of the vortex velocity resulting in a flux-flow instability \cite{Lar86inb,Bez92pcs,Shk17prb}. This instability becomes apparent as a sudden transition of the sample into the normal state. In the inset to Fig. \ref{f3}(b) we do not observe an abrupt jump, but rather a steep crossover from the flux-flow regime to the normally conducting state. This is because the flux-flow instability jump is expected to vanish above $0.4H_{c2}(0.8T_c) \thickapprox100$\,mT \cite{Lar86inb}. Nevertheless, the linewidth broadening in Fig. \ref{f4}(a) and the upturn in the CVC in Fig. \ref{f3}(b) in regime (III) are indicative of the diminishing long-range order of the vortex lattice that can explain the vanishing peak at $v>300$\,m/s.

To summarize, by using a spectrum analyzer connected to a nanowire meander near the surface of a superconducting Nb film we have detected an rf voltage induced by coherent vortex motion. The voltage is peaked at the washboard frequency and its subharmonics associated with the times needed for a vortex row to cross one meander period and the entire antenna, respectively. By sweeping the dc current value, we have been able to tune the generation frequency from $100$\,MHz to $1.5$\,GHz, thereby demonstrating that patterned hybrid nanostructures can be used as dc-tunable rf generators. The generation vanishes below the depinning transition, at high vortex velocities, and if the magnetic field value is tuned away from the matching configuration. Taken together, these findings underline the decisive role of the coherence in the vortex dynamics for the rf generation. Since the employed method relies on magnetic induction measurements rather than voltage, we anticipate that it can be applied to skyrmion systems as well \cite{Dia17prb}.

\vspace{0.5cm}
OD acknowledges the German Research Foundation (DFG) for support through Grant No 374052683 (DO1511/3-1). This work was supported by the European Cooperation in Science and Technology via COST Action CA16218 (NANOCOHYBRI). Further, funding from the European Commission in the framework of the program Marie Sklodowska-Curie Actions --- Research and Innovation Staff Exchange (MSCA-RISE) under Grant Agreement No. 644348 (MagIC) is acknowledged.
\vfill

%\bibliography{D:/bibliobase/fluxonics}
%\bibliographystyle{aipnum4-1}


\end{document}